\documentclass[aps,pra,showpacs,twocolumn,superscriptaddress]{revtex4}
\usepackage{amssymb}
\usepackage{amsmath}
\usepackage{graphicx}
\usepackage{epsfig}

\begin{document}

\title{Non-Hermitian description of the dynamics of inter-chain pair
tunnelling}
\author{X. Z. Zhang}
\affiliation{College of Physics and Materials Science, Tianjin Normal University, Tianjin
300387, China}
\author{L. Jin}
\affiliation{School of Physics, Nankai University, Tianjin 300071, China}
\author{Z. Song}
\email{songtc@nankai.edu.cn}
\affiliation{School of Physics, Nankai University, Tianjin 300071, China}

\begin{abstract}
We study inter-chain pair tunnelling dynamics based on an exact two-particle
solution for a two-leg ladder. We show that the Hermitian Hamiltonian shares
a common two-particle eigenstate with a corresponding non-Hermitian Hubbard
Hamiltonian in which the non-Hermiticity arises from an on-site interaction
of imaginary strength. Our results provides that the dynamic processes of
two-particle collision and across-legs tunnelling are well described by the
effective non-Hermitian Hubbard Hamiltonian based on the eigenstate
equivalence. We also find that any common eigenstate is always associated
with the emergence of spectral singularity in the non-Hermitian Hubbard
model. This result is valid for both Bose and Fermi systems and provides a
clear physical implication of the non-Hermitian Hubbard model.
\end{abstract}

\maketitle


\section{Introduction}

\label{sec_intro} Complex parameter in a Hamiltonian, such as imaginary
potential, has been investigated under the framework of non-Hermitian
quantum mechanics \cite{Klaiman1,Znojil,Makris,Musslimani,Bender
08,Jentschura,Fan,A.M38,A.M391,A.M392,ZXZ,LGR}. The usefulness of the
complex parameter can be explored by establishing a correspondence between a
non-Hermitian system and a real physical system in an analytically exact
manner. The discovery of a parity-time ($\mathcal{PT}$)\ symmetric
non-Hermitian Hamiltonian having an entirely real quantum-mechanical energy
spectrum \cite{Bender 98} stimulated the efforts of establishing $\mathcal{PT%
}$ symmetric quantum theory as a complex extension of conventional quantum
mechanics \cite{A.M,A.M36,Jones,Bender 99,Dorey 01,A.M43}. This
complex extension has profound theoretical and methodological implications
in many other subjects, ranging from quantum field theory and mathematical
physics \cite{ZnojilPLA,Brody,JonesJPA,Bender2007}, to solid state \cite%
{Bendix,West} and atomic physics \cite%
{Graefe1,Graefe2,Graefe3,Graefe5}.

One way of extracting the physical meaning of a pseudo-Hermitian Hamiltonian
with a real spectrum is to seek its Hermitian counterparts \cite%
{A.M38,A.M391,A.M392}. There exists another Hermitian Hamiltonian that
shares the complete or partial spectrum when the spectrum of a
pseudo-Hermitian Hamiltonian is real. The metric-operator theory outlined in
Ref. \cite{A.M} provides a mapping between a pseudo-Hermitian Hamiltonian
and an equivalent Hermitian counterpart. However, the obtained equivalent
Hermitian Hamiltonian is usually quite complicated \cite{A.M,JLPT}, and it
is difficult to determine whether it describes real physics or is just an
unrealistic mathematical object. An alternative way to establish the
connection between a pseudo-Hermitian Hamiltonian and a physical system is
considering the equivalence of eigenstates \cite%
{JLpseudo1,JLpseudo2,JLpseudo3}. A Hermitian scattering center at resonant
transmission shares the same wave function with the corresponding
non-Hermitian tight-binding lattice consisting of the Hermitian scattering
center with two additional $\mathcal{PT}$-symmetric on-site complex
potentials.

In this paper, we extend this approach to interacting particle systems. In
condensed matter physics, inter-chain (inter-layer) pair tunnelling is a
popular process, and is an important component for the mechanism of
superconductivity \cite{WheatleyPRB,WheatleyNature}. We consider a two-leg
system with inter-chain pair tunnelling. Based on the exact two-particle
solution, we show that if the two-particle dynamics mainly refers to a
specific invariant subspace, then the corresponding two-particle dynamics
can be described by an effective non-Hermitian Hubbard system with an
imaginary on-site interaction. For a given initial state, the strength of
the imaginary on-site interaction is determined by the relative velocity of
the two particles. When we consider the two-particle dynamics associated
with the probability gain in one leg of the Hermitian system, a set of
corresponding non-Hermitian Hamiltonians are related to the spectral
singularities. Therefore, the dynamical correspondence is sensitive to the
selection of the initial state. The particle-creation dynamics can be
realized by considering the time-reversal process of it, which corresponds
to the annihilation of two particles.\textbf{\ }On the other hand, the
two-particle tunnelling associated with decrease of the probability in the
other leg can be well described by a non-Hermitian Hubbard model with the
definite pair dissipation. Especially, when the relative group velocity
matches the strength of pair tunnelling, the two-particle probability will
exhibit a completely transfer from one leg to the other, which corresponds
to pair annihilation in the effective non-Hermitian system. From this point
of view, we unveil the connection between the interacting Hermitian and
non-Hermitian systems in the context of wavepacket dynamics.

This paper is organized as follows. In Sec. \ref{sec_model} we introduce the
model Hamiltonians and their symmetry. In Sec. \ref{sec_solution}, we
present the equivalence between the Hermitian Hamiltonian with inter-chain
pair tunnelling and non-Hermitian Hamiltonians with an imaginary on site
interaction. Sec. \ref{sec_dynamics} and Sec. \ref{sec_NHD} are devoted to
construct the connection between two types of the systems through wavepacket
dynamics. Section \ref{sec_summary} provides the summary and discussion.

\section{Model Hamiltonians}

\label{sec_model}We address a physically meaningful non-Hermitian
Hamiltonian by associating pair tunnelling with an imaginary on-site
interaction in a non-Hermitian Hubbard model. As an illustration, we
consider two simple models described by a Hermitian and a non-Hermitian
Hamiltonian.

The Hermitian Hamiltonian can be written as follows
\begin{equation}
H=H_{A}+H_{B}+H_{AB},  \label{Her}
\end{equation}%
and%
\begin{eqnarray}
H_{\rho } &=&-\kappa \sum_{j=1}^{N}\left( a_{\rho ,j}^{\dagger }a_{\rho
,j+1}+\text{H.c.}\right) \text{, }(\rho =A,B), \\
H_{AB} &=&-\frac{J}{2}\sum_{j=1}^{N}\left( a_{A,j}^{\dagger
}a_{A,j}^{\dagger }a_{B,j}a_{B,j}+\text{H.c.}\right) .
\end{eqnarray}%
Obviously, it represents a tight-binding system consisting of a two-leg
ladder, with each leg $H_{\rho }$ $\left( \rho =A,B\right) $ having
dimension $N$. The two legs are coupled through a pair tunnelling term $%
H_{AB}$, which operates on the motion of multi particles. The Hamiltonian
possesses two symmetries. One is the\ $\mathcal{P}$\ symmetry: here $%
\mathcal{P}$\ represents the space-reflection operator (or parity operator),
and the effect of the parity operator is $\mathcal{P}a_{A,j}^{\dagger }%
\mathcal{P}^{-1}=$\ $a_{B,j}^{\dagger }$. The other is the particle-number
symmetry, which ensures probability conservation and leads to the following
commutation relation
\begin{equation}
\lbrack \widehat{N}_{\rho },H]\neq 0,\text{but }[\sum_{\rho }\widehat{N}%
_{\rho },H]=0,
\end{equation}%
where $\widehat{N}_{\rho }=\sum_{i}a_{\rho ,i}^{\dagger }a_{\rho ,i}$ $%
\left( \rho =A,B\right) $ are the particle-number operators for the upper
and lower legs, respectively. The probability is conserved in the entire
system $H$, but breaks in subsystems $H_{A}$\ and $H_{B}$. The inter-chain
pair tunnelling admits a peculiar symmetry,\textbf{\ }%
\begin{equation}
\lbrack \left( -1\right) ^{\widehat{N}_{\rho }},H]=0,  \label{inter_symmetry}
\end{equation}%
i.e., the conservation of particle-number parity. %

\begin{figure}[tbp]
\centering
\includegraphics[ bb=80 290 550 750, width=0.45\textwidth, clip]{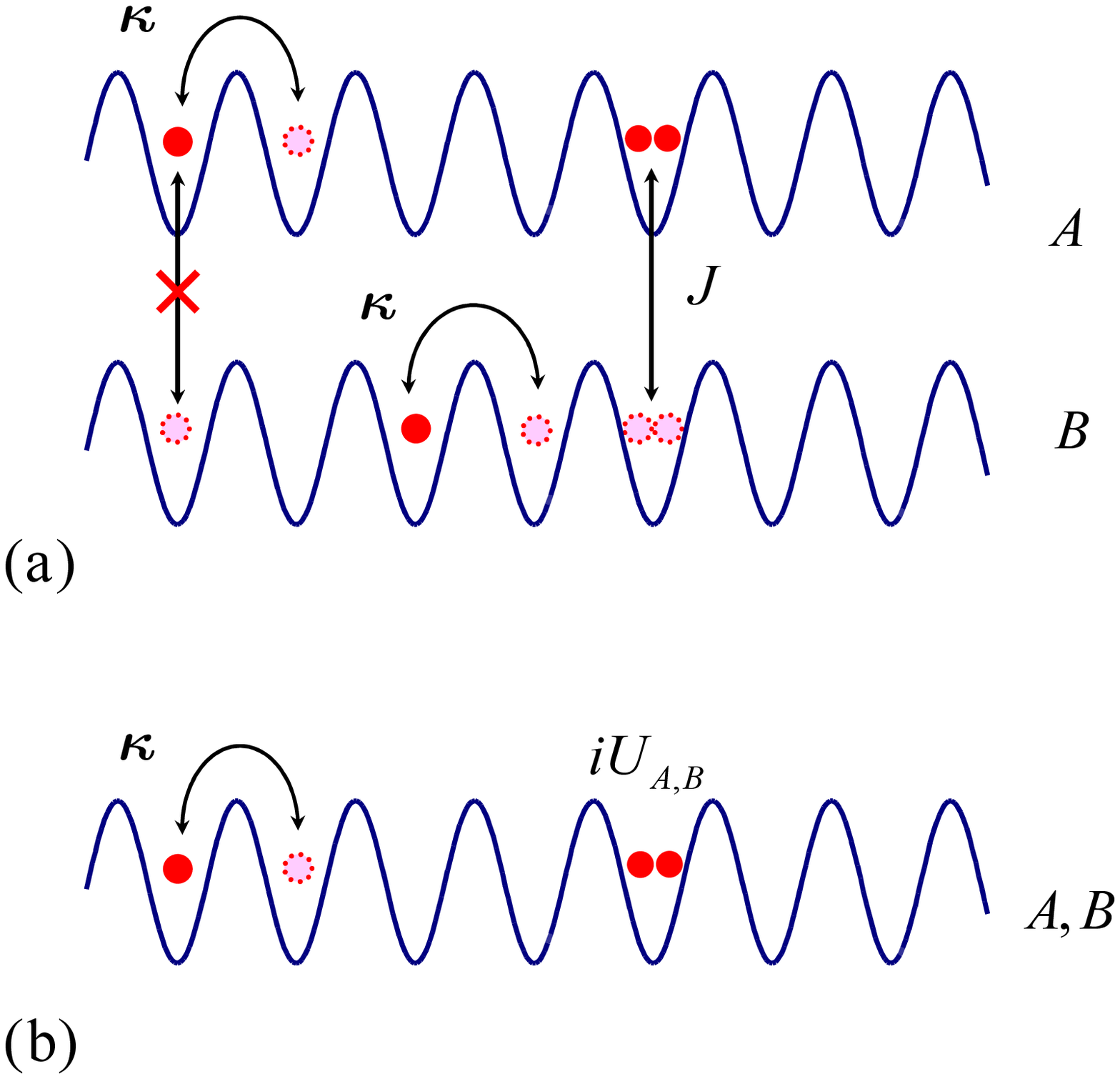}
\caption{(Color online) Schematic illustration of the concerned lattice
systems. (a) Two-leg ladder for non-interacting particles. Two particles at
the same site can hop simultaneously across two legs with $J$ being the
inter-chain pair tunnelling strength. (b) Two independent Hubbard chains
with imaginary on-site interaction $iU_{\protect\rho }$$\left( \protect\rho %
=A,B\right)$.}
\label{fig1}
\end{figure}

Another related system is a non-Hermitian system composed by two independent
Hubbard chains, which can be expressed as%
\begin{equation}
\mathcal{H}=\mathcal{H}_{A}+\mathcal{H}_{B},  \label{nonHer}
\end{equation}%
and%
\begin{equation}
\mathcal{H}_{\rho }=-\kappa \sum_{i=1}^{N}\left( a_{\rho ,i}^{\dagger
}a_{\rho ,i+1}+\text{H.c.}\right) +\frac{iU_{\rho }}{2}\sum_{i}n_{\rho
,i}\left( n_{\rho ,i}-1\right) ,
\end{equation}%
where $\rho =A,B$. The non-Hermiticity of $\mathcal{H}_{\rho }$ arises from
the complex on-site interaction $iU_{\rho }$.

We note that $\mathcal{H}$\ has the same symmetries as $H$ does, i.e., $%
\left[ \mathcal{H}_{\rho },\sum_{i}n_{\rho ,i}\right] =0$, $[\mathcal{H}%
,\sum_{\rho ,i}n_{\rho ,i}]=0$, except $\left[ \mathcal{H}_{A},\mathcal{H}%
_{B}\right] =0$. This allows us to construct the eigenstates of two models
in the same invariant subspaces. For instance, particle-preserving symmetry
leads to the two-particle invariant subspace, which can be further
decomposed into two invariant subspaces with basis sets $\{a_{A,i}^{\dagger
}a_{B,j}^{\dagger }\left\vert 0\right\rangle \}$\ and $\{a_{\rho
,i}^{\dagger }a_{\rho ,j}^{\dagger }\left\vert 0\right\rangle \}$,
respectively. In the next section, we will investigate the connection
between the two-particle solutions of these two Hamiltonians. In Fig. \ref%
{fig1}, we schematically illustrate the system $H$ and $\mathcal{H}$.

\section{Pair tunnelling and spectral singularity}

\label{sec_solution}Now we turn to study the two-particle eigenstates of $H$
and $\mathcal{H}$, from which we expect to establish the connection between
two models. We focus on the solutions in the invariant subspace spanned by $%
\{a_{\rho ,i}^{\dagger }a_{\rho ,j}^{\dagger }\left\vert 0\right\rangle \}$,
i.e., both particles are either in chain $A$ or $B$. The derivation in
Appendix \ref{sec_App_1} shows that for each given $\{K,k\}$ with $K\in %
\left[ -\pi ,\pi \right] ,k\in \left[ 0,\pi \right] $, there are two
degenerate eigenstates of $H$\ with energy\textbf{\ }%
\begin{equation}
\varepsilon _{K}\left( k\right) =-4\kappa \cos \left( K/2\right) \cos k.
\end{equation}%
And the associated eigenstates can be written as%
\begin{equation}
\left\vert \psi _{K,k}^{\pm }\right\rangle =\sum_{r\geqslant 0,\rho
=A,B}f_{K,k}^{\rho ,\pm }\left( r\right) \left\vert \phi _{r}^{\rho }\left(
K\right) \right\rangle ,  \label{eigen state}
\end{equation}%
\ and
\begin{eqnarray}
\left\vert \phi _{0}^{\rho }\left( K\right) \right\rangle &=&\frac{1}{2\sqrt{%
N}}\sum_{j}e^{iKj}a_{\rho ,j}^{\dagger }a_{\rho ,j}^{\dagger }\left\vert
\text{vac}\right\rangle , \\
\left\vert \phi _{r}^{\rho }\left( K\right) \right\rangle &=&\frac{e^{iKr/2}%
}{\sqrt{N}}\sum_{j}e^{iKj}a_{\rho ,j}^{\dagger }a_{\rho ,j+r}^{\dagger
}\left\vert \text{vac}\right\rangle \text{, }  \notag \\
&&\left( r>1\right) ,
\end{eqnarray}%
where $\left\vert \phi _{0}^{\rho }\left( K\right) \right\rangle $ and $%
\left\vert \phi _{r}^{\rho }\left( K\right) \right\rangle $ are
translational invariant bases. The corresponding wavefunctions $%
f_{K,k}^{\rho ,\pm }\left( r\right) $ can be expressed explicitly as
\begin{eqnarray}
f_{K,k}^{A,+}\left( r\right) &=&f_{K,k}^{B,-}\left( r\right)  \notag \\
&=&\left\{
\begin{array}{cc}
e^{-ikr}+\eta _{K,k}e^{ikr}, & r>0 \\
\left( 1+\eta _{K,k}\right) /\sqrt{2}, & r=0%
\end{array}%
\right. , \\
f_{K,k}^{B,+}\left( r\right) &=&f_{K,k}^{A,-}\left( r\right)  \notag \\
&=&\left\{
\begin{array}{cc}
\xi _{K,k}e^{ikr}, & r>0 \\
\left( 1+\xi _{K,k}\right) /\sqrt{2}, & r=0%
\end{array}%
\right. ,
\end{eqnarray}%
where%
\begin{eqnarray}
\eta _{K,k} &=&\frac{\lambda _{K,k}^{2}-J^{2}}{\lambda _{K,k}^{2}+J^{2}},%
\text{ }\xi _{K,k}=-\frac{2i\lambda _{K,k}J}{\lambda _{K,k}^{2}+J^{2}}, \\
\lambda _{K,k} &=&4\kappa \cos \left( K/2\right) \sin k.
\end{eqnarray}%
We note that $K$ represents the central momentum vector of two particles,
while $k$ represents the relative momentum between the two particles. In
this sense, the eigenstates $\left\vert \psi _{K,k}^{\pm }\right\rangle $
are associated with the dynamic process in which two particles collide with
each other in one leg and then tunnel into the other leg.

\begin{figure}[tbp]
\centering
\includegraphics[bb=50 190 540 757, width=0.47\textwidth, clip]{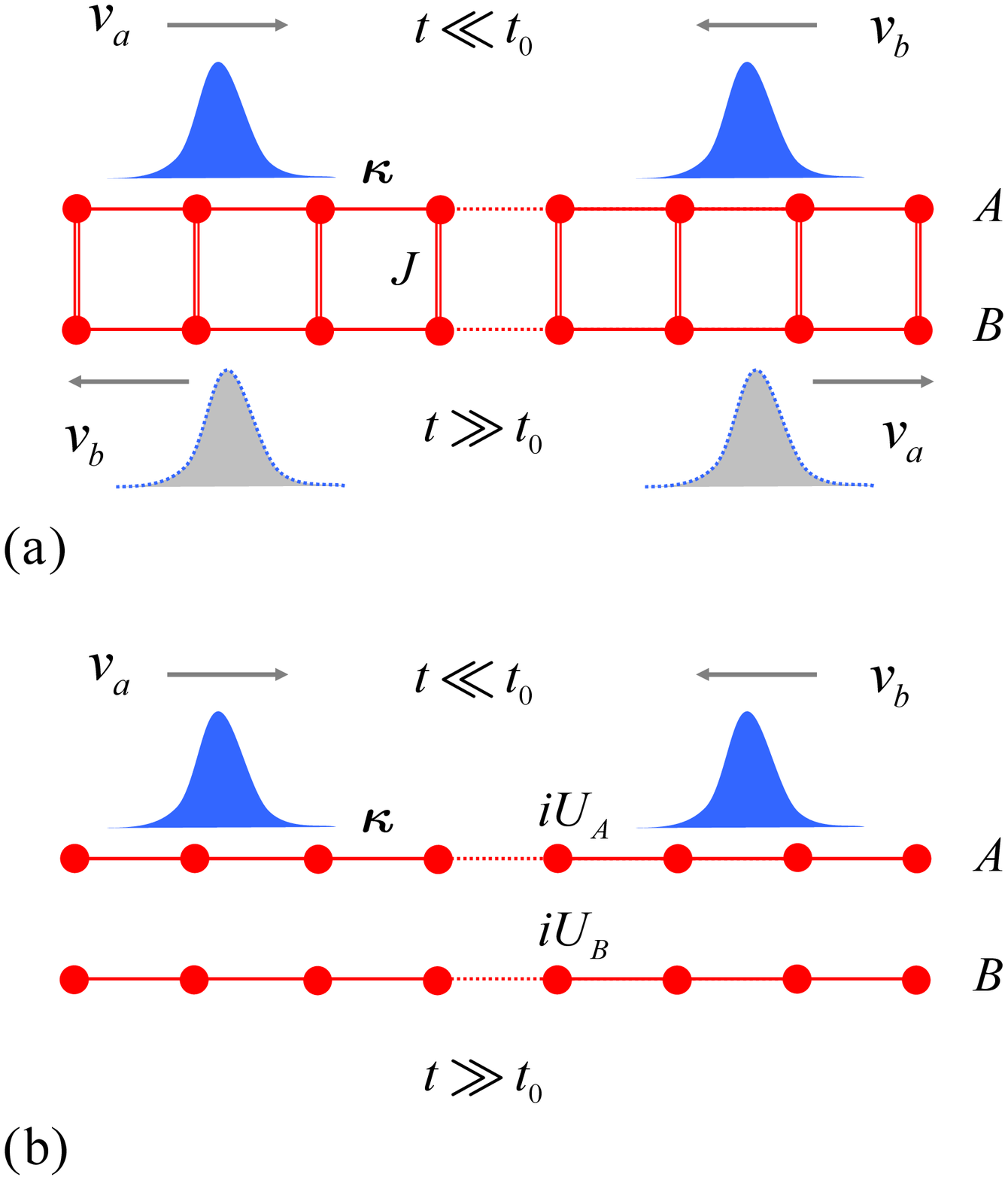}
\caption{(Color online) Schematic illustration of the dynamics of two
separable wavepackets placed initially at leg $A$. (a) The two wavepackets
enter into leg $B$ completely associated with the probability flow from leg $%
A$ into leg $B$ when $J=\protect\upsilon _{r}$. The dynamic process of two
such particles in leg $A $ can be described approximately through the
non-Hermitian interacting system in (b) with $U_{A}=\protect\upsilon _{r}$ ($%
\protect\upsilon _{r}<0$), in which the annihilation of the two wavepackets
occurs through the imaginary on-site interaction.}
\label{fig2}
\end{figure}

Similarly, we can construct the eigenstates of $\mathcal{H}$ having the same
form in Eq. (\ref{eigen state}) based on the result shown in Appendix \ref%
{sec_App_2},
\begin{equation}
\left\vert \chi _{K,k}^{\pm }\right\rangle =\sum_{r\geqslant 0,\rho
=A,B}g_{K,k}^{\rho ,\pm }\left( r\right) \left\vert \phi _{r}^{\rho }\left(
K\right) \right\rangle ,
\end{equation}%
where
\begin{eqnarray}
g_{K,k}^{A,+}\left( r\right) &=&g_{K,k}^{A,-}\left( r\right)  \notag \\
&=&\left\{
\begin{array}{cc}
e^{-ikr}+\mu _{K,k}e^{ikr}, & r>0 \\
\left( 1+\mu _{K,k}\right) /\sqrt{2}, & r=0%
\end{array}%
\right. ,  \label{ga} \\
g_{K,k}^{B,+}\left( r\right) &=&-g_{K,k}^{B,-}\left( r\right)  \notag \\
&=&\left\{
\begin{array}{cc}
e^{-ikr}+\nu _{K,k}e^{ikr}, & r>0 \\
\left( 1+\nu _{K,k}\right) /\sqrt{2}, & r=0%
\end{array}%
\right. ,  \label{gb}
\end{eqnarray}%
and the parameters are%
\begin{equation}
\mu _{K,k}=\frac{\lambda _{K,k}+U_{A}}{\lambda _{K,k}-U_{A}},\text{ }\nu
_{K,k}=\frac{\lambda _{K,k}+U_{B}}{\lambda _{K,k}-U_{B}}.
\end{equation}%
It is easy to check that when the following conditions are satisfied%
\begin{equation}
U_{A}=-J^{2}/\lambda _{K,k},U_{B}=\lambda _{K,k},
\end{equation}%
we could obtain
\begin{equation}
\left\vert \psi _{K,k}^{+}\right\rangle =\left\vert \chi
_{K,k}^{+}\right\rangle .  \label{cr1}
\end{equation}%
Note that the eigenstates $\left\vert \chi _{K,k}^{\pm }\right\rangle $ are
the functions of $U_{A}$ and $U_{B}$. The equivalence condition (\ref{cr1})
denotes that the $U_{A}$ and $U_{B}$ are $\left\{ K,k\right\} $ dependent.
Thus one requires two indices to label the eigenstate as $\left\vert \chi
_{K,k}^{\pm }\left( U_{A},U_{B}\right) \right\rangle $. For the sake of
convenience, we neglect the $\left( U_{A},U_{B}\right) $ of $\left\vert \chi
_{K,k}^{\pm }\left( U_{A},U_{B}\right) \right\rangle $. If we exchange the
values of $U_{A}$ and $U_{B}$
\begin{equation}
U_{A}=\lambda _{K,k},U_{B}=-J^{2}/\lambda _{K,k},
\end{equation}%
we have%
\begin{equation}
\left\vert \psi _{K,k}^{-}\right\rangle =\left\vert \chi
_{K,k}^{+}\right\rangle ,  \label{cr2}
\end{equation}%
which arises from the parity symmetry of both $H$\ and $\mathcal{H}$. This
indicates that the two Hamiltonians have common eigenstates, revealing the
connection between a Hermitian and a non-Hermitian Hamiltonian. This
connection has the following features: (i) We find that $iU_{A}$ and $iU_{B}$
are $\left\{ K,k\right\} $ dependent and for a given $\left\{ K,k\right\} $,
they are all imaginary but with different signs, representing a
complementarity pair gain and loss. Further investigation in the next
section will show that this ensures the conservation of particles in the%
\textbf{\ }whole system. (ii) As an independent non-Hermitian Hubbard chain
with on-site strength $iU_{\rho }$, the derivation in Appendix \ref%
{sec_App_3} shows that when$\ U_{\rho }=\lambda _{K,k}$ this Hamiltonian has
a spectral singularity at point $\left\{ K,k\right\} $.\ (iii) Furthermore,
we find that in the case of $J^{2}=\lambda _{K,k}^{2}$, two independent
non-Hermitian Hubbard chains have a spectral singularity simultaneously at
point $\left\{ K,k\right\} $. The mechanism of the occurrence of the
spectral singularity and the corresponding physical implications will be
addressed in the next section.

\section{Tunnelling dynamics}

\label{sec_dynamics}Considering two local particles in one of two legs,
which have no overlap with each other, the tunnelling term would have zero
effect on the dynamics. But when the two particles meet, particle transfer
occurs between two legs. The pair transmission probability depends on many
factors as discussed in the following. In this section, we will investigate
the dynamics of two-wavepackets collision based on the above formalism. We
start our investigation from the time evolution of an initial state as%
\begin{equation}
\left\vert \Phi \left( 0\right) \right\rangle =\left\vert \Phi
_{A,a}\right\rangle \left\vert \Phi _{A,b}\right\rangle ,  \label{Phi}
\end{equation}%
which represents two separable boson wavepackets $a$ and $b$. Here%
\begin{equation}
\left\vert \Phi _{\rho ,\gamma }\right\rangle =\frac{1}{\sqrt{\Omega }}%
\sum_{j}e^{-\alpha ^{2}\left( j-N_{\gamma }\right) ^{2}}e^{ik_{\gamma
}j}a_{\rho ,j}^{\dagger }\left\vert \text{Vac}\right\rangle ,
\end{equation}%
with $\gamma =a$, $b$, and $\rho =A$, $B$\ represents a Gaussian wavepacket,
which has a width $2\sqrt{\ln 2}/\alpha $, a central position $N_{\gamma }$
in chain $\rho $\ and a group velocity $\upsilon _{\gamma }=-2\kappa \sin
k_{\gamma }$. The condition that $N_{a}-N_{b}\gg 1/\alpha $ ensures that two
initial bosons cannot overlap, and thus having no pair tunnelling.
Straightforward derivation shows that%
\begin{eqnarray}
\left\vert \Phi \left( 0\right) \right\rangle &=&\frac{1}{2}\sum_{\sigma
=\pm }\left( \left\vert \Phi _{A,a}\right\rangle \left\vert \Phi
_{A,b}\right\rangle +\sigma \left\vert \Phi _{B,a}\right\rangle \left\vert
\Phi _{B,b}\right\rangle \right)  \notag \\
&=&\frac{1}{\sqrt{2\Omega _{1}}}\sum_{K}e^{-\left( K-2k_{c}\right)
^{2}/4\alpha ^{2}}  \notag \\
&&\times e^{-iN_{c}\left( K-2k_{c}\right) }\left\vert \psi _{K}^{\pm }\left(
r_{c},q_{c}\right) \right\rangle ,  \label{psi}
\end{eqnarray}%
where%
\begin{equation}
\left\vert \psi _{K}^{\pm }\left( r_{c},q_{c}\right) \right\rangle =\frac{1}{%
\sqrt{\Omega _{2}}}\sum_{r}e^{-\alpha ^{2}\left( r-r_{c}\right)
^{2}/2}e^{iq_{c}r/2}\left\vert \phi _{r}^{\pm }\left( K\right) \right\rangle
,
\end{equation}%
and $\Omega _{1,2}$ is the normalized factor. Here we have used the
following transformations%
\begin{eqnarray}
N_{c} &=&\frac{1}{2}\left( N_{a}+N_{b}\right) ,r_{c}=N_{b}-N_{a}, \\
\ k_{c} &=&\frac{1}{2}\left( k_{a}+k_{b}\right) ,q_{c}=k_{b}-k_{a}, \\
l &=&j+r,
\end{eqnarray}%
and identities
\begin{eqnarray}
&&2\left[ \left( j-N_{a}\right) ^{2}+\left( l-N_{b}\right) ^{2}\right]
\notag \\
&=&\left[ \left( j+l\right) -\left( N_{a}+N_{b}\right) \right] ^{2}  \notag
\\
&&+\left[ \left( l-j\right) -\left( N_{b}-N_{a}\right) \right] ^{2}, \\
&&2\left( k_{a}j+k_{b}l\right)  \notag \\
&=&\left( k_{a}+k_{b}\right) \left( j+l\right) +\left( k_{b}-k_{a}\right)
\left( l-j\right) .
\end{eqnarray}%
%
%
%
%
%
%
%
%
%
%
%
%
%
%
%
%
%
%
%
%
%
%

\begin{figure}[tbp]
\centering
\includegraphics[ bb=21 364 577 672, width=0.45\textwidth, clip]{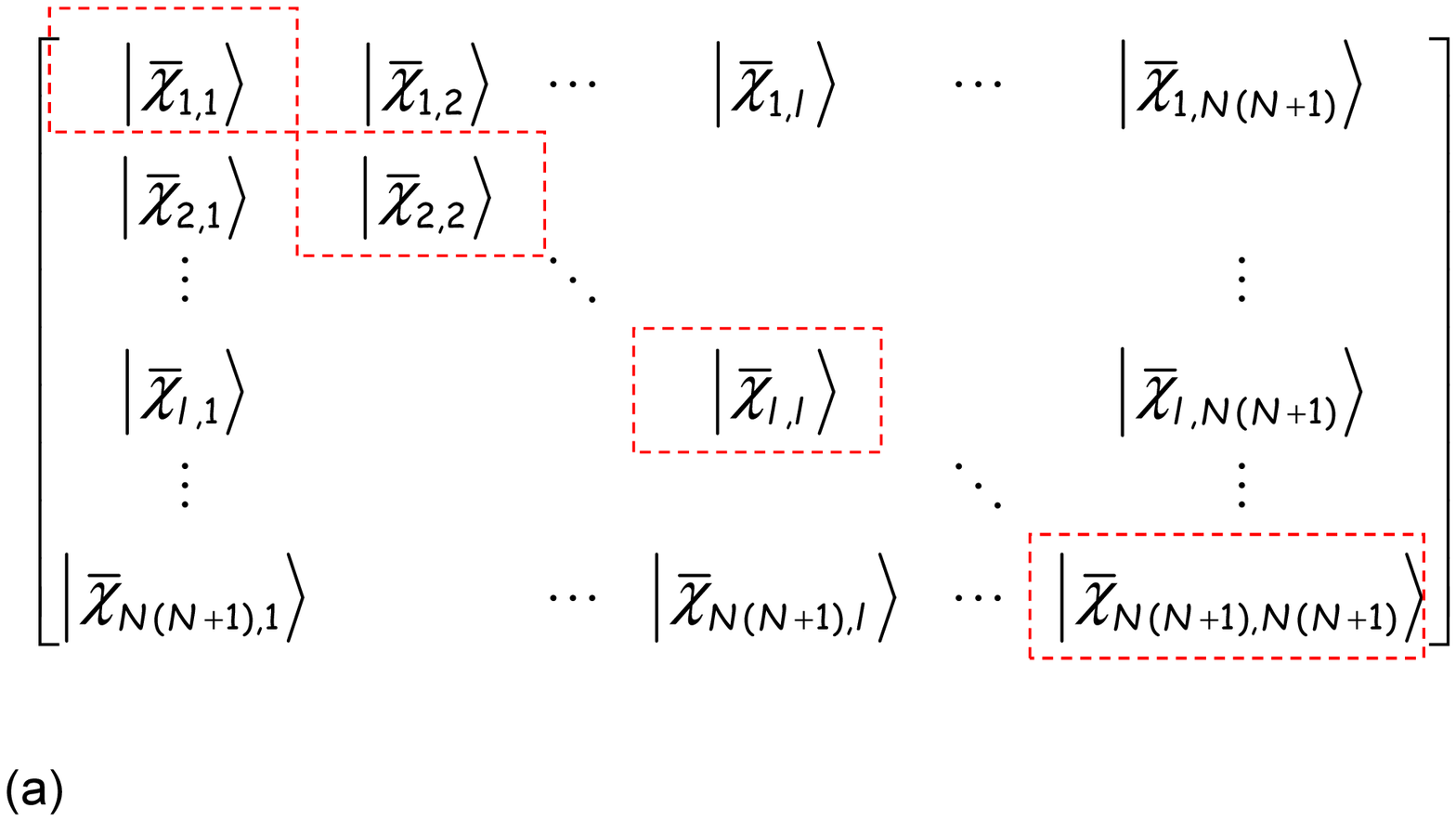} %
\vskip 0.5 cm
\includegraphics[ bb=21 364 577 672, width=0.45\textwidth,
clip]{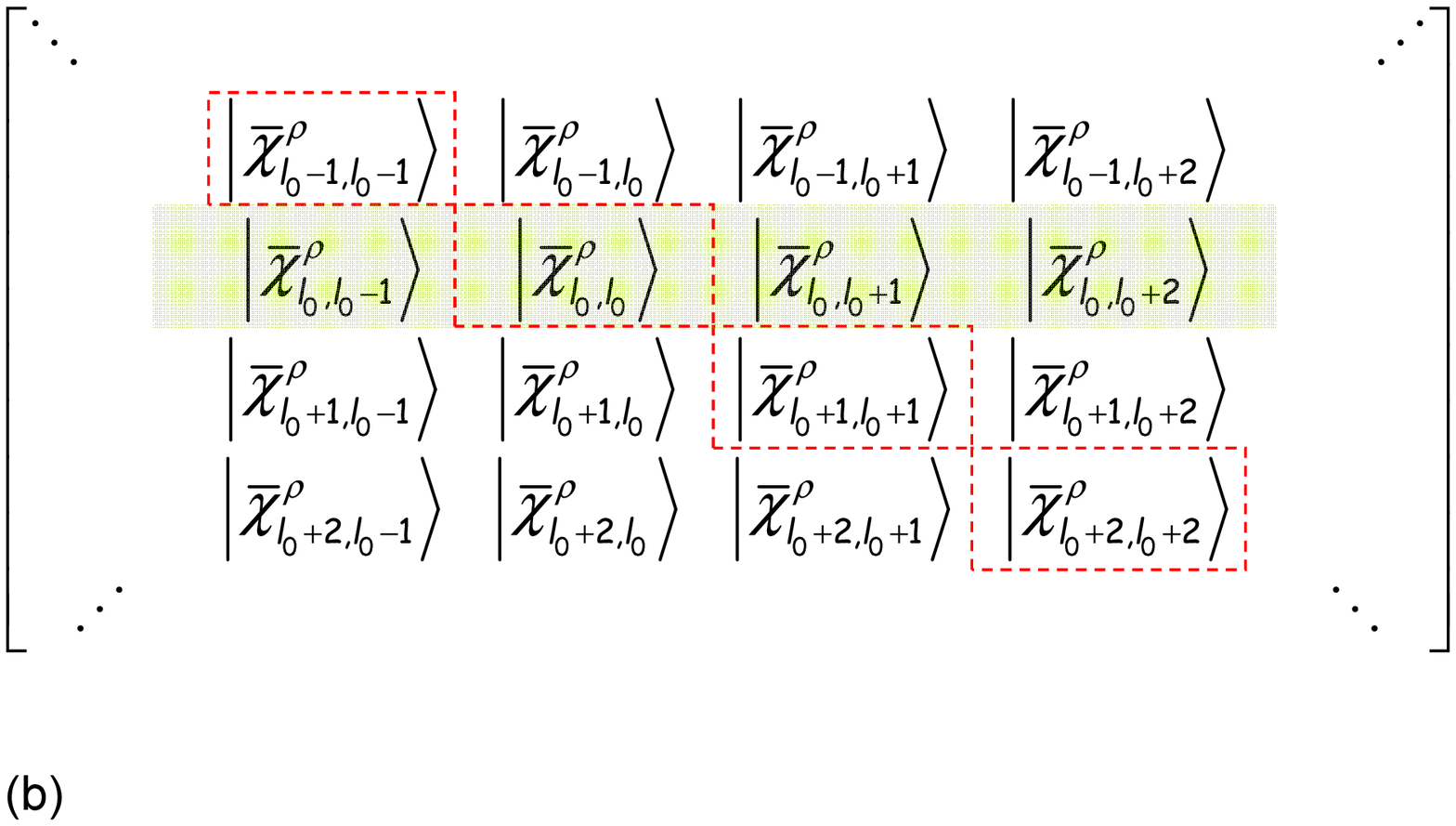} 
\par
\caption{(Color online) The ket matrix to illustrate the connection between
eigenstates of $\overline{\mathcal{H}}_{l}$ and $H$. (a) States in lth row \{%
$\left\vert \bar{\protect\chi}_{l,l^{\prime }}\right\rangle $\} represents
complete set of eigenstates of $\overline{\mathcal{H}}_{l}$. The diagonal
states \{$\left\vert \bar{\protect\chi}_{l,l}\right\rangle $\} (dashed box)
is the complete set of eigenstates of $H$. (b) A block around certain ket $%
\left\vert \bar{\protect\chi}_{l_{0},l_{0}}^{\protect\rho}\right\rangle $
satisfying the Eqs. (\protect\ref{c1}), (\protect\ref{c2}), (\protect\ref{c3}%
) and (\protect\ref{c4}). All the rows in such a block are identical
approximately. Then the diagonal states can be replaced by that in a row
(green shadowed).}
\label{fig3}
\end{figure}
%
%
%
%
%
%

We note that the component of state $\left\vert \Phi \left( 0\right)
\right\rangle $\ on each invariant subspace represents an incident
wavepacket along the chains described by\ $H_{\text{eq}}^{K,\pm }$ with a
width $2\sqrt{\ln 2}/\alpha $, a central position $r_{c}=N_{b}-N_{a}$ and a
group velocity $\upsilon =-4\kappa \cos \left( K/2\right) \sin \left(
q_{c}/2\right) $. It is worth pointing out that as $\alpha \ll 1$, the
initial state is distributed mainly in the invariant subspace $K=2k_{c}$,
where the wavepacket moves with the group velocity\textbf{\ }$\upsilon _{r}=$
$-4\kappa \cos \left( k_{c}\right) \sin \left( q_{c}/2\right) =$ $\upsilon
_{b}-\upsilon _{a}$. Then the time evolution of state $\left\vert \Phi
\left( t\right) \right\rangle $\ can be derived by the evolution of
wavepacket in two chains $H_{\text{eq}}^{K,\pm }$, which eventually can be
obtained from the solution in Eq. (\ref{fkj}).\ Furthermore, according to
the solution, the evolved state of $\left\vert \psi _{K}^{\pm }\left(
r_{c},q_{c}\right) \right\rangle $ can be expressed approximately in the form%
\textbf{\ }of $e^{i\beta \left( r_{c}^{\prime }\right)
}R_{2k_{c},q_{c}/2}^{\pm }\left\vert \psi _{K}^{\pm }\left( r_{c}^{\prime
},-q_{c}\right) \right\rangle $, which represents a reflected wavepacket in
the equivalent semi-infinite chain $H_{\text{eq}}^{K,\pm }$. The expressions
of $R_{2k_{c},q_{c}/2}^{\pm }$\ and\ $H_{\text{eq}}^{K,\pm }$\ are given in
the Appendix \ref{sec_App_2}. Here $\beta \left( r_{c}^{\prime }\right) $,
as a function of the position of the reflected wavepacket, is an overall
phase and is independent of $J$. We assume that the collision occurs at
instant $t_{0}$, the evolved state at time $t\gg t_{0}$ has the form of%
\begin{eqnarray}
\left\vert \Phi \left( t\right) \right\rangle &=&\sum_{\sigma =\pm }\Omega
^{-1}e^{i\beta \left( \left\vert N_{a}^{\prime }-N_{b}^{\prime }\right\vert
\right) }R_{2k_{c},q_{c}/2}^{\sigma }  \notag \\
&&\times \sum_{j,l}e^{-\alpha ^{2}\left( l-N_{b}^{\prime }\right)
^{2}}e^{-\alpha ^{2}\left( j-N_{a}^{\prime }\right) ^{2}}  \notag \\
&&\times e^{ik_{b}j}e^{ik_{a}l}\left( a_{A,j}^{\dagger }a_{A,l}^{\dagger
}+\sigma a_{B,j}^{\dagger }a_{B,l}^{\dagger }\right) \left\vert \text{Vac}%
\right\rangle .  \label{psit}
\end{eqnarray}%
which also represents two separable wavepackets at $N_{a}^{\prime }$\ and $%
N_{b}^{\prime }$, respectively. Comparing Eqs. (\ref{psi}) and (\ref{psit}),
it is straightforward to figure out that the two-particle wavepackets behave
as classical particles, which swap the momenta with each other after
collision. For simplicity, we denote an incident single-particle wavepacket
as $\left\vert \lambda ,p,A\right\rangle $, where $\lambda =\mathrm{L,}$ $%
\mathrm{R}$\ indicates the particle coming from the left or right of the
collision zone, and $p$ is the central momentum. In this context, we give
the asymptotic expression for the collision process\textbf{\ }in the
following: at time $t\ll t_{0}$, we have%
\begin{equation}
\left\vert \mathrm{L},k_{a},A\right\rangle \left\vert \mathrm{R}%
,k_{b},A\right\rangle =\frac{1}{\sqrt{2}}\left( \left\vert \mathrm{F}%
^{+}\right\rangle +\left\vert \mathrm{F}^{-}\right\rangle \right) ,
\end{equation}%
where%
\begin{equation}
\left\vert \mathrm{F}^{\pm }\right\rangle =\frac{1}{\sqrt{2}}\left(
\left\vert \mathrm{L},k_{a},A\right\rangle \left\vert \mathrm{R}%
,k_{b},A\right\rangle \pm \left\vert \mathrm{L},k_{a},B\right\rangle
\left\vert \mathrm{R},k_{b},B\right\rangle \right) .
\end{equation}%
and after collision, at time $t\gg t_{0}$, the wavepackets exchange their
momenta, which admits
\begin{eqnarray}
&&\left\vert \mathrm{L},k_{a},A\right\rangle \left\vert \mathrm{R}%
,k_{b},A\right\rangle \pm \left\vert \mathrm{L},k_{a},B\right\rangle
\left\vert \mathrm{R},k_{b},B\right\rangle  \notag \\
&\longmapsto &R_{2k_{c},q_{c}/2}^{\pm }\left( \left\vert \mathrm{L}%
,k_{b},A\right\rangle \left\vert \mathrm{R},k_{a},A\right\rangle \right.
\notag \\
&&\left. \pm \left\vert \mathrm{L},k_{b},B\right\rangle \left\vert \mathrm{R}%
,k_{a},B\right\rangle \right) .
\end{eqnarray}%
By neglecting the $J$-independent overall phase, therefore we have
\begin{eqnarray}
&&\left\vert \mathrm{L},k_{a},A\right\rangle \left\vert \mathrm{R}%
,k_{b},A\right\rangle  \notag \\
&\longmapsto &\cos \Delta _{2k_{c},q_{c}/2}\left\vert \mathrm{L}%
,k_{b},A\right\rangle \left\vert \mathrm{R},k_{a},A\right\rangle  \notag \\
&&+i\sin \Delta _{2k_{c},q_{c}/2}\left\vert \mathrm{L},k_{b},B\right\rangle
\left\vert \mathrm{R},k_{a},B\right\rangle ,  \label{wp collision 1}
\end{eqnarray}%
where
\begin{eqnarray}
R_{2k_{c},q_{c}/2}^{\pm } &=&e^{\pm i\Delta _{2k_{c},q_{c}/2}},\text{ } \\
\Delta _{2k_{c},q_{c}/2} &=&2\tan ^{-1}\left( -\frac{J}{\lambda
_{2k_{c},q_{c}/2}}\right) ,
\end{eqnarray}%
as discussed in Appendix \ref{sec_App_1}. Evidently, Eq. (\ref{wp collision
1}) shows that after collision, one part of two wavepackets in leg $A$,
which corresponds to the first term in Eq. (\ref{wp collision 1}), is
reflected as two identical classical particles. Meanwhile another part,
which corresponds to the second term in Eq. (\ref{wp collision 1}), tunnels
into leg $B$.

Considering a special case with $\upsilon _{r}=J$, i.e., the pair-tunnelling
amplitude is equal to the relative group velocity, we have $\Delta
_{2k_{c},q_{c}/2}=\pi /2$, and this leads to
\begin{equation}
\left\vert \mathrm{L},k_{a},A\right\rangle \left\vert \mathrm{R}%
,k_{b},A\right\rangle \longmapsto i\left\vert \mathrm{L},k_{b},B\right%
\rangle \left\vert \mathrm{R},k_{a},B\right\rangle .
\end{equation}%
Clearly, this represents the process that two separable wavepackets on leg $%
A $\ tunnel into leg $B$\ completely.

\section{Non-Hermitian dynamics}

\label{sec_NHD}

From the above discussions regarding the dynamics of across-leg tunneling,
we see that the two-particle probability transfers from one leg to another.
The two-particle probability in one leg is not conserved. Thus, a natural
question to ask is whether there exists an effective non-Hermitian
Hamiltonian for characterizing such a dynamics. To this end, we first
present the connections between Hermitian Hamiltonian $H$ and $\mathcal{H}$
in a compact form. There are $N\left( N+1\right) $ eigenstates of $H$ in the
invariant subspace spanned by $\{a_{\rho ,i}^{\dag }a_{\rho ,j}^{\dag
}\left\vert Vac\right\rangle \}$. Each of the eigenstates $\left\{
\left\vert \psi _{K,k}^{\pm }\right\rangle \right\} $ corresponds to a
specific eigenstate $\left\vert \chi _{K,k}^{+}\right\rangle $ of the
non-Hermitian Hubbard chain with the $\left( K,k\right) $-dependent
interaction $iU_{\rho }$ as in Eqs. (\ref{cr1}) and (\ref{cr2}). We note
that the eigenstates of $H$ and $\left( K,k\right) $-dependent Hamiltonian $%
\mathcal{H}$ are related to the index $\left( K,k\right) $. In the
following, we take a single index $\eta $\ to represent $\left( K,k\right) $%
. For the system with $2N$ sites, all possible $\left( K,k\right) $\ is
denoted as $\eta =1,2,...,N(N+1)/2$.\ The eigenstates of $H$ is denoted as $%
\left\vert \bar{\psi}_{l}\right\rangle $ ($l\in \lbrack 1,N(N+1)]$) with%
\begin{eqnarray}
&&\left\vert \bar{\psi}_{\eta }\right\rangle \equiv \left\vert \psi
_{K,k}^{+}\right\rangle ,  \notag \\
&&\left\vert \bar{\psi}_{\eta +N(N+1)/2}\right\rangle \equiv \left\vert \psi
_{K,k}^{-}\right\rangle .
\end{eqnarray}%
Accordingly, the $\left( K,k\right) $-dependent Hamiltonian $\mathcal{H}$ is
denoted as $\overline{\mathcal{H}}_{l}$ with%
\begin{eqnarray}
&&\overline{\mathcal{H}}_{\eta }\equiv \mathcal{H}\left( K,k\right) ,\text{ }
\notag \\
&&\text{for }U_{A}=-J^{2}/\lambda _{K,k},U_{B}=\lambda _{K,k}, \\
&&\overline{\mathcal{H}}_{\eta +N(N+1)/2}\equiv \mathcal{H}\left( K,k\right)
,\text{ }  \notag \\
&&\text{for }U_{A}=\lambda _{K,k},U_{B}=-J^{2}/\lambda _{K,k}.
\end{eqnarray}%
\begin{widetext}
The eigenstate of $\overline{\mathcal{H}}_{l}$\ is denoted as $\left\vert
\bar{\chi}_{l,l^{\prime }}\right\rangle $ with%
\begin{eqnarray}
&&\left\vert \bar{\chi}_{\eta ,\eta ^{\prime }}\right\rangle \equiv
\left\vert \chi _{K^{\prime },k^{\prime }}^{+}\right\rangle ,\text{ }%
\left\vert \bar{\chi}_{\eta ,\eta ^{\prime }+N(N+1)/2}\right\rangle \equiv
\left\vert \chi _{K^{\prime },k^{\prime }}^{-}\right\rangle ,  \notag \\
&&\text{for }\mathcal{H}\left( K,k\right) \text{ with }U_{A}=-J^{2}/\lambda
_{K,k},\text{ }U_{B}=\lambda _{K,k}, \\
&&\left\vert \bar{\chi}_{\eta +N(N+1)/2,\eta ^{\prime }}\right\rangle \equiv
\left\vert \chi _{K^{\prime },k^{\prime }}^{-}\right\rangle ,\text{ }%
\left\vert \bar{\chi}_{\eta +N(N+1)/2,\eta ^{\prime }+N(N+1)/2}\right\rangle
\equiv \left\vert \chi _{K^{\prime },k^{\prime }}^{+}\right\rangle ,  \notag
\\
&&\text{for }\mathcal{H}\left( K,k\right) \text{ with }U_{A}=\lambda _{K,k},%
\text{ }U_{B}=-J^{2}/\lambda _{K,k}.
\end{eqnarray}%
\end{widetext}Note that the eigenstate $\left\vert \bar{\chi}_{l,l^{\prime
}}\right\rangle $ possesses two subscripts. The first one indicates the $%
\left( K,k\right) $-dependent on-site interactions $U_{A}$ and $U_{B}$, and
the second one denotes the center and relative momenta $\left( K,k\right) $
of the eigenstate for a given $U_{A}$ and $U_{B}$. In Fig. \ref{fig3}(a), we
illustrate the $\left\vert \bar{\chi}_{l,l^{\prime }}\right\rangle $ via ket
matrix. The states in $l$th row represents the complete set of eigenstates
of $\overline{\mathcal{H}}_{l}$. Based on this notation, the Schrodinger
equations become compact%
\begin{eqnarray}
H\left\vert \bar{\psi}_{l}\right\rangle &=&E_{l}\left\vert \bar{\psi}%
_{l}\right\rangle , \\
\overline{\mathcal{H}}_{l}\left\vert \bar{\chi}_{l,l^{\prime }}\right\rangle
&=&\varepsilon _{l,l^{\prime }}\left\vert \bar{\chi}_{l,l^{\prime
}}\right\rangle .
\end{eqnarray}%
Note that $\varepsilon _{l,l^{\prime }}$\ is related to the scattering
solution of two particles, which possesses the form of $\varepsilon
_{l,l^{\prime }}=-4\kappa \cos (K^{\prime }/2)\cos k^{\prime }$, where $%
\left( K^{\prime },k^{\prime }\right) $\ denotes possible center and
relative momentuma. These eigenstates have simple relations%
\begin{equation}
\left\vert \bar{\psi}_{l}\right\rangle =\left\vert \bar{\chi}%
_{l,l}\right\rangle ,\text{ }E_{l}=\varepsilon _{l,l}=-4\kappa \cos
(K/2)\cos k,
\end{equation}%
which indicate that the diagonal states $\left\{ \left\vert \bar{\chi}%
_{l,l}\right\rangle \right\} $ of Fig. \ref{fig3}(a) is the complete set of
eigenstates of $H$. Here, $\left\vert \bar{\psi}_{\eta }\right\rangle $ ($%
\left\vert \bar{\psi}_{\eta +N(N+1)/2}\right\rangle $) represents that the
two particles collide with each other in leg $A$ ($B$) and then tunnel into
leg $B$ ($A$).

\begin{figure*}[tbp]
\centering
\includegraphics[bb=0 205 518 600, width=0.45\textwidth, clip]{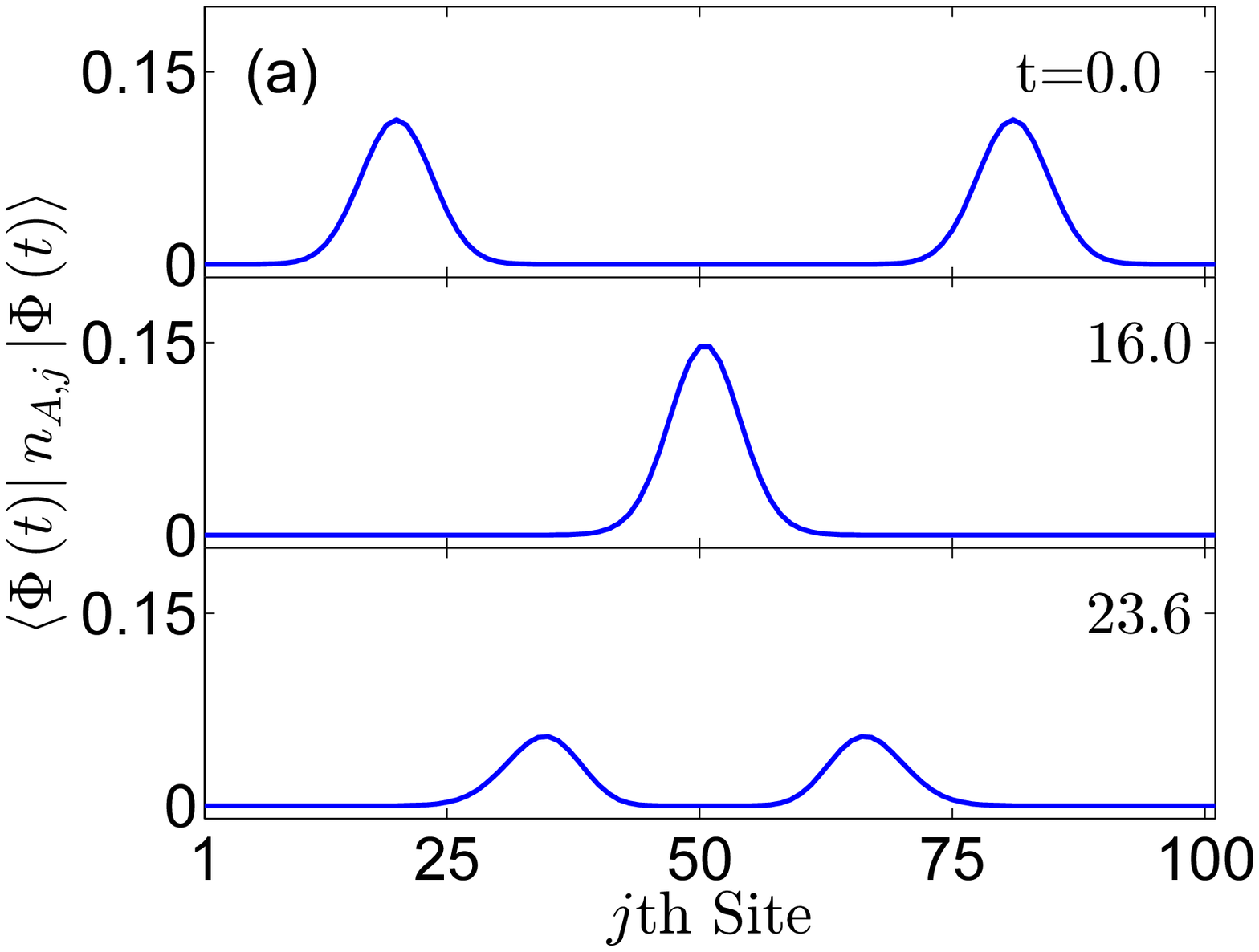} %
\includegraphics[ bb=0 205 518 600, width=0.45\textwidth, clip]{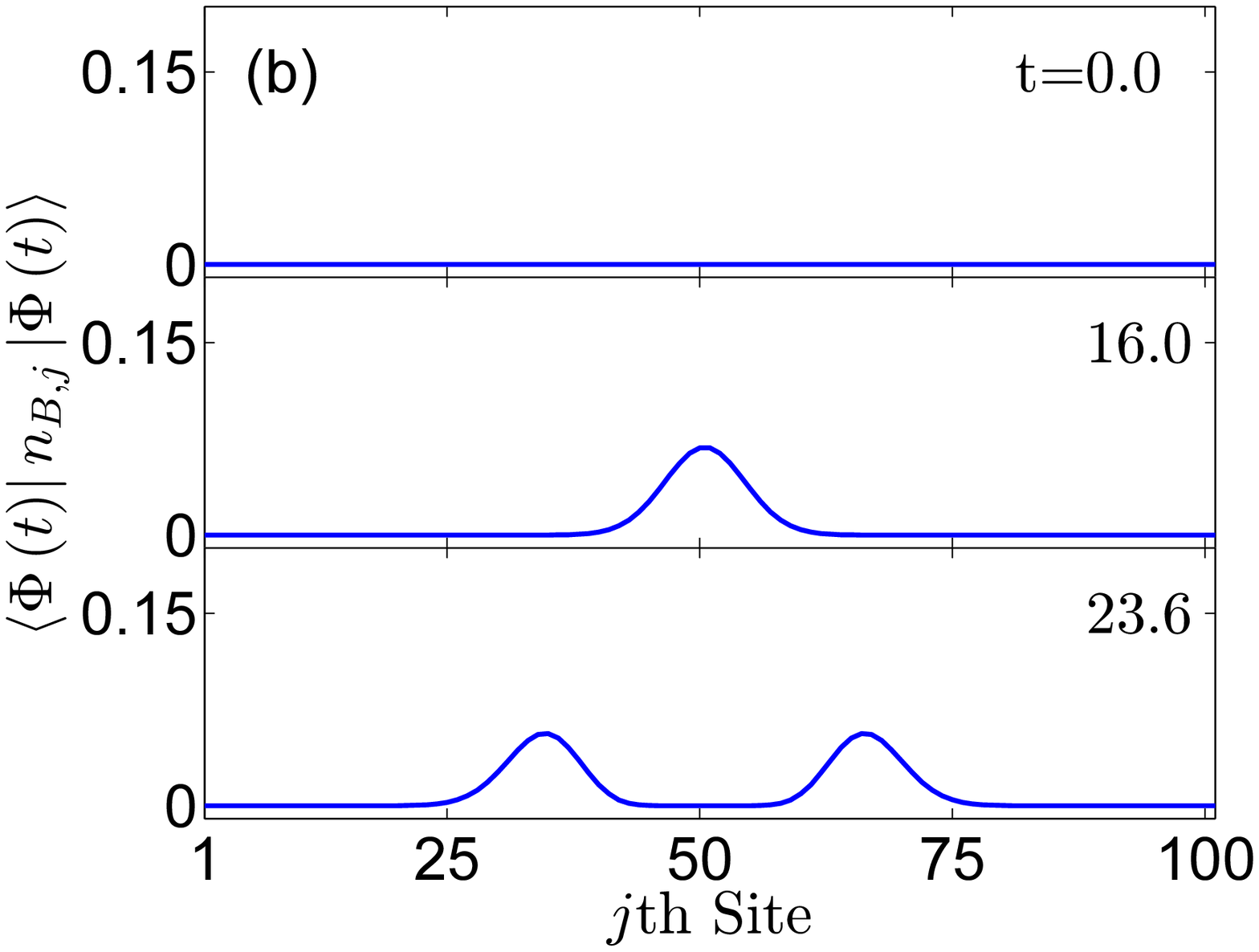} %
\includegraphics[ bb=0 205 518 600, width=0.45\textwidth, clip]{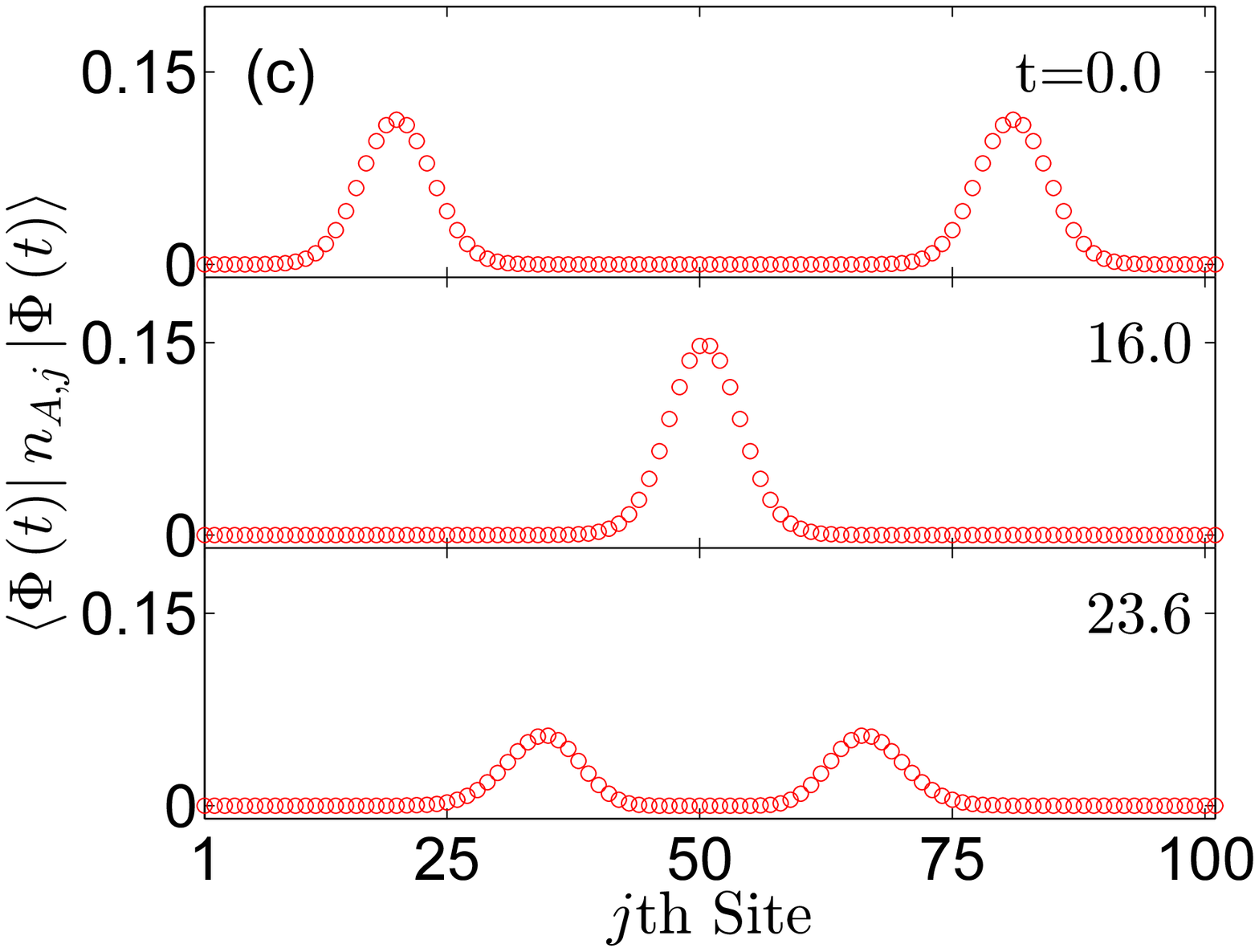} %
\includegraphics[ bb=0 205 518 600, width=0.45\textwidth, clip]{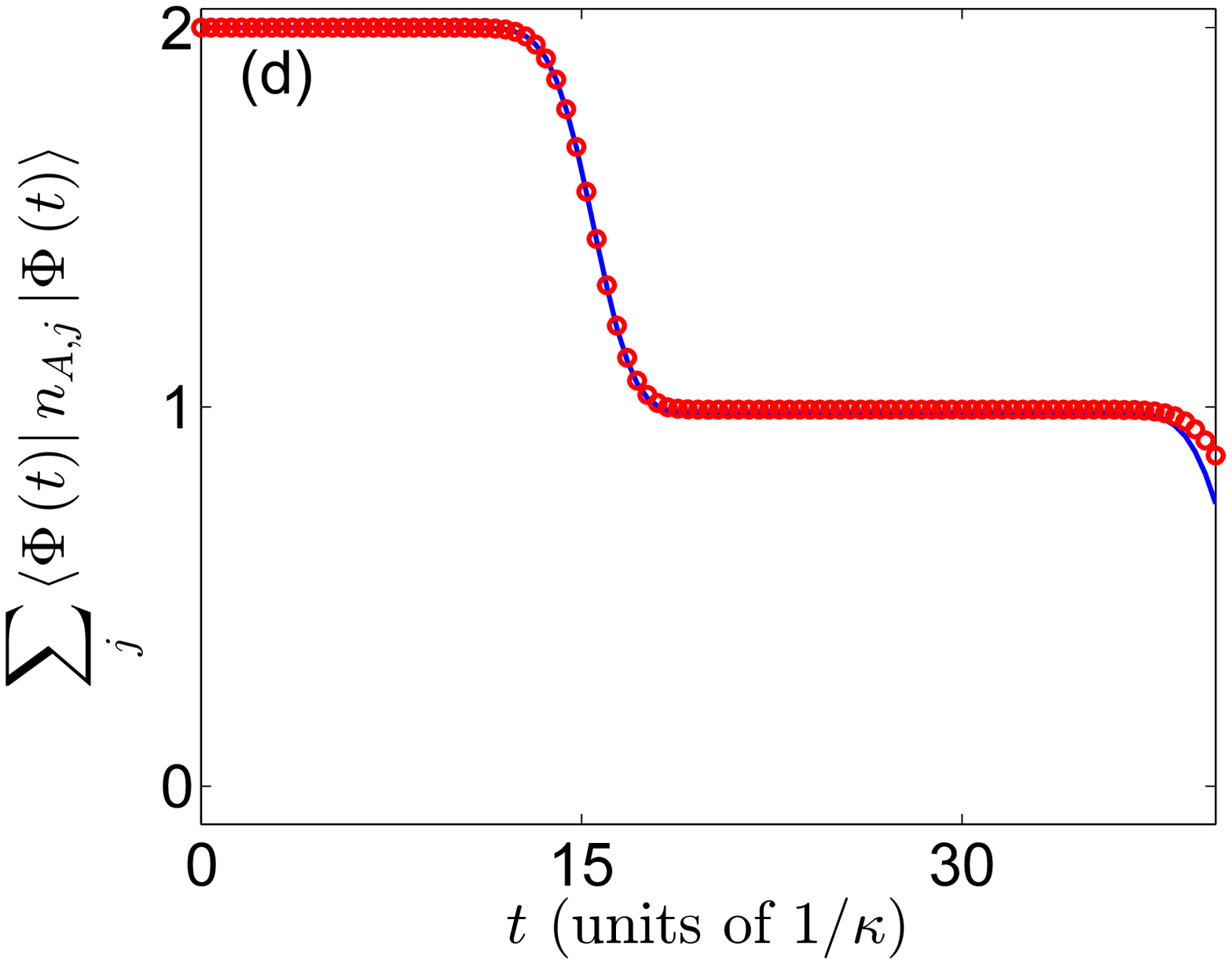}
\caption{(Color online)Probability distribution of the evolved wave function
for the initial state being two-particle incident Gaussian wave packets with
$k_{a}=\protect\pi /2$, $N_{a}=20$, $k_{b}=-\protect\pi /2$, $N_{a}=80$ in
leg $A$ of different systems. (a) and (b) depict the probability $\left\vert
\left\langle \Phi \left( t\right) \right\vert n_{\protect\rho ,j}\left\vert
\Phi \left( t\right) \right\rangle \right\vert ^{2}$ ($\protect\rho =A,$ $B$%
) of two-leg Hermitian system $H$ for leg $A $ and $B$ with a tunnelling
strength $J=4\left( 3-2\protect\sqrt{2}\right) $, respectively. (c) denotes
the corresponding probability distribution on leg $A$ of the non-Hermitian
system $\mathcal{H}$ with imaginary on-site interaction $U_{A}=-4\left( 3-2%
\protect\sqrt{2}\right) ^{2} $. (d) The red circle and blue line represent
the total probability $\sum_{j}\left\vert \left\langle \Phi \left( t\right)
\right\vert n_{A,j}\left\vert \Phi \left( t\right) \right\rangle \right\vert
^{2}$ of leg $A$ as functions of time $t$ for the Hermitian system $H$ and
the non-Hermitian system $\mathcal{H}$, respectively. We plot $\left\vert
\left\langle \Phi \left( t\right) \right\vert n_{j}\left\vert \Phi \left(
t\right) \right\rangle \right\vert ^{2}$ at different instant $t$ in units
of ($1/\protect\kappa $). One can see that when the matching condition $%
U_{A}=-J^{2}/\protect\upsilon _{r}$ is satisfied, the non-Hermitian
Hamiltonian can be utilized to describe the dynamics of leg $A$ of Hermitian
Hamiltonian $H$, which is in accordance with the theoretical analysis in the
text.}
\label{fig4}
\end{figure*}

When considering the dynamical correspondence in the non-Hermitian Hubbard
system, there exists two kinds of dynamical processes corresponding to $%
\left\vert \bar{\psi}_{\eta }\right\rangle $ and $\left\vert \bar{\psi}%
_{\eta +N(N+1)/2}\right\rangle $: (i) $\left\vert \bar{\chi}_{\eta ,\eta
}^{A}\right\rangle \equiv \left( \left\vert \bar{\chi}_{\eta ,\eta
}\right\rangle +\left\vert \bar{\chi}_{\eta ,\eta +N(N+1)/2}\right\rangle
\right) /\sqrt{2}$ denotes the two-particle collision process in leg $A$
accompanied by the decrease of the two-particle probability while $%
\left\vert \bar{\chi}_{\eta ,\eta }^{B}\right\rangle \equiv \left(
\left\vert \bar{\chi}_{\eta ,\eta }\right\rangle -\left\vert \bar{\chi}%
_{\eta ,\eta +N(N+1)/2}\right\rangle \right) /\sqrt{2}$ represents a process
related to the increase of two-particle probability in leg $B$. (ii) $%
\left\vert \bar{\chi}_{\eta +N(N+1)/2,\eta }^{B}\right\rangle \equiv \left(
\left\vert \bar{\chi}_{\eta +N(N+1)/2,\eta }\right\rangle -\left\vert \bar{%
\chi}_{\eta +N(N+1)/2,\eta +N(N+1)/2}\right\rangle \right) /\sqrt{2}$
represents the two-particle collision process in leg $B$ accompanied by the
decrease of the two-particle probability, and $\left\vert \bar{\chi}_{\eta
+N(N+1)/2,\eta }^{A}\right\rangle \equiv \left( \left\vert \bar{\chi}_{\eta
+N(N+1)/2,\eta }\right\rangle +\left\vert \bar{\chi}_{\eta +N(N+1)/2,\eta
+N(N+1)/2}\right\rangle \right) /\sqrt{2}$ denotes a process associated with
the increase of two-particle probability in leg $A$. For a collision process
along leg $\rho $ $\left( \rho =A,B\right) $ in Hermitian systems, there are
$N\left( N+1\right) /2$ related non-Hermitian Hamiltonians. Therefore one
cannot obtain a Hubbard chain with a certain value of $iU_{\rho }$ to
describe the dynamics along one of two legs. However, for an initial state
distributed mainly in the vicinity of $\left\vert \bar{\psi}%
_{l_{0}}\right\rangle $, the correspondence of dynamics can be characterized
by the eigenstates around $l_{0}$th row in which the value of $l_{0}$ is
determined by the central and relative momenta $\left( K_{0},k_{0}\right) $
of the considered initial state. This corresponds to a block around certain
ket $\left\vert \bar{\chi}_{l_{0},l_{0}}\right\rangle $, which can be shown
in Fig. \ref{fig3}(b). For the sake of simplicity and convenience, we
confine the discussions to the case of $l_{0}\in \left[ 1,\left( N+1\right)
N/2\right] $. The conclusion still holds for the case of $l_{0}\in \left[
\left( N+1\right) N/2+1,\left( N+1\right) N\right] $, in which the $U_{A}$
and $U_{B}$ exchange their values. To seek an effective non-Hermitian
Hamiltonian to characterize such a dynamics, we first consider the collision
dynamics in leg $A$, which is accompanied by the decrease of two-particle
probability. If the involved wavefunctions changes slowly around $\left(
\left\vert \bar{\chi}_{l_{0},l_{0}}\right\rangle +\left\vert \bar{\chi}%
_{l_{0},l_{0}+N\left( N+1\right) /2}\right\rangle \right) /\sqrt{2}$, then
one can use an effective non-Hermitian Hamiltonian $\mathcal{H}_{A}\left(
K_{0},k_{0}\right) $ with a definite $U_{A}=-J^{2}/\lambda _{K_{0},k_{0}}$
as an approximation to describe such a dynamics of leg $A$ in the Hermitian
system. To this end, we take the derivative of the function $\mu _{K,k}$
with respect to $K$\ and $k$,%
\begin{eqnarray}
\left( \frac{\partial \mu _{K,k}}{\partial K}\right) _{K_{0},k_{0}}
&=&-\Lambda \sin \left( K_{0}/2\right) \sin k_{0},  \label{c1} \\
\left( \frac{\partial \mu _{K,k}}{\partial k}\right) _{K_{0},k_{0}}
&=&2\Lambda \cos \left( K_{0}/2\right) \cos k_{0},  \label{c2}
\end{eqnarray}%
where $\Lambda =8\kappa J^{2}\lambda _{K_{0},k_{0}}/\left( \lambda
_{K_{0},k_{0}}^{2}+J^{2}\right) ^{2}$. The optimal condition can be achieved
when\textbf{\ }$\left( \partial \mu _{K,k}/\partial k\right) _{K_{0},k_{0}}=0
$, and $\left( \partial \mu _{K,k}/\partial K\right) _{K_{0},k_{0}}=0$. This
can be realized through adjusting the relative group velocity of the initial
two wavepackets. The condition also indicates that all the rows in such a
block are identical approximately as shown in Fig. \ref{fig3}(b). Then the
diagonal states of the block can be replaced by that in a row with green
shadowed. On the other hand, for the dynamics along leg $B$, each of the
eigenstates in the vicinity of $\left\vert \bar{\chi}_{l_{0},l_{0}}^{B}%
\right\rangle =\left( \left\vert \bar{\chi}_{l_{0},l_{0}}\right\rangle
-\left\vert \bar{\chi}_{l_{0},l_{0}+N\left( N+1\right) /2}\right\rangle
\right) /\sqrt{2}$\ corresponds to a spectral singularity of the
non-Hermitian Hamiltonian $\mathcal{H}_{B}\left( K,k\right) $. This leads to
the coefficients\textbf{\ }$\lambda _{K,k}=U_{B}$ $\left( \nu _{K,k}=\infty
\right) $. In order to avoid this divergence, we can rewrite the expression
of Eq. (\ref{gb}) in the form
\begin{eqnarray}
g_{K,k}^{B,+}\left( r\right)  &=&-g_{K,k}^{B,-}\left( r\right)   \notag \\
&=&\left\{
\begin{array}{cc}
\varsigma _{k}e^{-ikr}+\zeta _{k}e^{ikr}, & r>0 \\
2\lambda _{K,k}/\sqrt{2}, & r=0%
\end{array}%
\right. .  \label{gb2}
\end{eqnarray}%
where $\varsigma _{k}=\lambda _{K,k}-U_{B}$, and $\zeta _{k}=\lambda
_{K,k}+U_{B}$. Here we want to point out that the relative magnitude between
the amplitudes of right-going wave $e^{ikr}$ and left-going wave $e^{-ikr}$
is meaningful, since we focus on the scattering solution in the limit of $%
N\rightarrow \infty $. In this sense, the form of the wavefunctions $%
g_{K,k}^{A,\sigma }\left( r\right) $ and $g_{K,k}^{B,\sigma }\left( r\right)
$ $\left( \sigma =\pm \right) $ are not unique. After multiplying $K-k$
dependent constant, the renormalized scattering solutions are still the
corresponding eigenstates of $\mathcal{H}_{A}$ and $\mathcal{H}_{B}$. In the
definition of Eq. (\ref{gb}), the existence of the spectral singularity in system can be
determined by either $\mu _{K,k}=0$ $\left( \nu _{K,k}=0\right) $ or $\mu
_{K,k}=\infty $ $\left( \nu _{K,k}=\infty \right) $, which is associated
with the pair-annihilation or pair-creation process. This corresponds to the
case of $\zeta _{k}=0$ or $\varsigma _{k}=0$ in Eq. (\ref{gb2}). To obtain
the effective non-Hermitian Hamiltonian, we focus on the variation of $%
\varsigma _{k}$\ in the vicinity of $\left\vert \bar{\chi}%
_{l_{0},l_{0}}^{B}\right\rangle $\ in the ket matrix. As we have done in leg
$A$, we take the partial derivative of the $\varsigma _{k}$\ with respect to
$K$\ and $k$, respectively, which yields
\begin{eqnarray}
\left( \frac{\partial \varsigma _{k}}{\partial K}\right) _{K_{0},k_{0}} &=&0,
\label{c3} \\
\left( \frac{\partial \varsigma _{k}}{\partial k}\right) _{K_{0},k_{0}} &=&0.
\label{c4}
\end{eqnarray}%
This indicates that one can replace the diagonal states with row states for
any given momentuma $\left( K_{0},k_{0}\right) $\ in the ket matrix as shown
in Fig. \ref{fig3}(b). Thus, an effective Hamiltonian with definite $%
U_{B}=\lambda _{K_{0},k_{0}}$\ can be employed to simulate the dynamics of
leg $B$\ in the Hermitian system. Here we want to stress that there is no
tunneling between non-Hermitian Hamiltonians $\mathcal{H}_{A}$ and $\mathcal{%
H}_{B}$. Thus, we cannot employ an effective non-Hermitian Hamiltonian $%
\mathcal{H}$ with definite strengths of the pair dissipation and gain to
describe the tunneling dynamics between two legs. The dynamical
correspondence of leg $B$\ can be obtained through another method, which
will be detailed in the following.

In parallel, we can investigate the dynamics of a two-wavepacket collision
by analyzing the time evolution of the initial state $\left\vert \Phi \left(
0\right) \right\rangle $\ in the effective non-Hermitian system\textbf{\ }$%
\mathcal{H}$\textbf{\ }with $U_{A}=-J^{2}/\upsilon _{r}$. Similarly, we can
obtain the asymptotic expression for the collision process as%
\begin{equation}
\left\vert \mathrm{L},k_{a},A\right\rangle \left\vert \mathrm{R}%
,k_{b},A\right\rangle \longmapsto \cos \Delta _{2k_{c},q_{c}/2}\left\vert
\mathrm{L},k_{b},A\right\rangle \left\vert \mathrm{R},k_{a},A\right\rangle ,
\label{wp collision 2}
\end{equation}%
which has the same form as the wave function in leg $A$\ of Eq. (\ref{wp
collision 1}). This indicates that the effective non-Hermitian Hamiltonian%
\textbf{\ }$\mathcal{H}$\textbf{\ }can describe the wavepacket dynamics in
subsystem (leg $A$) of a Hermitian system $H$. Naturally, when the strength
of the imaginary on-site interaction $U_{A}$ is equal to the relative group
velocity $\upsilon _{r}$ ($\Delta _{2k_{c},q_{c}/2}=\pi /2$), the two
particles will exhibit a behaviour of pair annihilation in leg $A$ and will
never tunnel into leg $B$. This process is schematically illustrated in Fig. %
\ref{fig2}(b). Note that $\mathcal{H}$ cannot describe the wavepacket
dynamics in leg $B$, because there is no tunnelling between leg $A$\ and $B$%
\ in\textbf{\ }$\mathcal{H}$. However, the final state $\left\vert \mathrm{L}%
,k_{b},B\right\rangle \left\vert \mathrm{R},k_{a},B\right\rangle $\ in leg $%
B $\ as in Eq. (\ref{wp collision 1}) can be prepared by using non-Hermitian
Hamiltonian $H_{B}$\ in another way. To this end, we require an initial
state to simulate the creation of a pair of particles. Moreover, the modulus
of the initial state should tend to be $0$, owing to the fact that no one
can create a pair of particles out of nothing. Then the initial state driven
by the $\mathcal{H}_{B}$ will evolve to\textbf{\ }$\left\vert \mathrm{L}%
,k_{b},B\right\rangle \left\vert \mathrm{R},k_{a},B\right\rangle $\textbf{\ }%
accompanied by the increase of two-particle probability. However the
selection of such an initial state is too cumbersome. There is a lot of
states with near-zero-modulus value. The different types of the initial
states will exhibit distinct dynamical behaviors. In other words, the
dynamics of the system is sensitive to the initial state. Therefore, the
elaborately selection of the initial state is a crucial step to successfully
mimic the dynamics of leg $B$\ in Hermitian system. Fortunately, we can
chose the initial state by considering the time-reversal process of the
dynamics of leg $B$, which corresponds to the annihilation of two
wavepackets $\left\vert \mathrm{L},k_{b},B\right\rangle \left\vert \mathrm{R}%
,k_{a},B\right\rangle $. This can be realized through adjusting the on-site
interaction $U_{B}$\ and relative group velocity $\upsilon _{r}$\ based on
the result obtained in leg $A$. In this sense, the final near-zero-modulus
state can be selected as an initial state of the particles creation process.
And the corresponding driven Hamiltonian can also be obtained by taking the
time-reversal operation on the related non-Hermitian Hubbard Hamiltonian
with pair annihilation.

In order to further validate the conclusion obtained above, we compare the
local-state dynamics in two such systems by numerical simulation. To do
this, we introduce the quantity $\left\vert \left\langle \Phi \left(
t\right) \right\vert n_{j}\left\vert \Phi \left( t\right) \right\rangle
\right\vert ^{2}$\ to characterize the shape and probability distribution of
the two wavepackets in Fig. \ref{fig4}. For the Hermitian case as shown in
Fig. \ref{fig4}(a) and \ref{fig4}(b), one can see that when the two
wavepackets enter into the interaction region, the probability of the two
wavepacket transfers from leg $A$\ to $B$\ due to the pair tunnelling $J$.
The process of the decrease of the two-wavepacket probability in leg $A$\
can also be approximately described through the two-wavepacket dynamics in
an effective non-Hermitian Hubbard Hamiltonian\textbf{\ }$\mathcal{H}$%
\textbf{\ }with $U_{A}=-J^{2}/\upsilon _{r}$, as is shown in Fig. \ref{fig4}%
(c).

\section{Summary}

\label{sec_summary} In summary, we have studied the inter-chain pair
tunnelling dynamics based on the exact two-particle solution of a two-leg
ladder. It is shown that the Hermitian Hamiltonian shares a common
two-particle eigenstate with a corresponding non-Hermitian Hubbard model, in
which the non-Hermiticity arises from an imaginary on-site interaction. Such
a common state is associated with the spectral singularity of the equivalent
non-Hermitian system. The dynamical correspondence is dependent on the
selection of the initial state. For the dynamics accompanied with the
increase of the two-particle probability, such an initial state can be
obtained through a time-reversal process of the annihilation of two
wavepackets. On the other hand, the reduction of the two-particle
probability in the other leg of the Hermitian system can be well
characterized by the effective non-Hermitian Hubbard model with the definite
strength of pair dissipation, which is also determined by the relative and
center momentuma of the initial state. In addition, we have also found that
the two particles display perfect transfer from one leg to the other when $%
\upsilon _{r}=J$, which corresponds to the pair annihilation in the
effective non-Hermitian Hubbard system with the strength of the imaginary
on-site interaction $\upsilon _{r}=U_{\rho }$. This result is valid for both
Bose and Fermi systems and provides a clear physical implication of the
non-Hermitian Hubbard model.

\section{Appendix}

\subsection{Solution of the\emph{\ }two-leg ladder}

\label{sec_App_1}In this section, we derive the solution of the Hamiltonian
shown in Eq. (\ref{Her}) in a two-particle invariant subspace. Here, we take
the periodic boundary condition that\textbf{\ }$a_{\rho ,j}=a_{\rho ,j+N}$.
Due to the symmetry in Eq. (\ref{Her}), which preserves the parity of
particle number in each leg, the $\mathcal{P}$ symmetry, and the
translational symmetry, the basis spanning the subspace can be constructed
as
\begin{eqnarray}
&&\left\vert \varphi _{0}^{\pm }\left( K,r\right) \right\rangle =\frac{1}{2%
\sqrt{N}}\sum_{j}e^{iKj}\left( a_{A,j}^{\dagger }a_{A,j}^{\dagger }\right.
\notag \\
&&\left. \pm a_{B,j}^{\dagger }a_{B,j}^{\dagger }\right) \left\vert \text{vac%
}\right\rangle , \\
&&\left\vert \varphi _{r}^{\pm }\left( K,r\right) \right\rangle =\frac{1}{%
\sqrt{2N}}e^{iKr/2}\sum_{j}e^{iKj}\left( a_{A,j}^{\dagger
}a_{A,j+r}^{\dagger }\right.  \notag \\
&&\text{ }\left. \pm a_{B,j}^{\dagger }a_{B,j+r}^{\dagger }\right)
\left\vert \text{vac}\right\rangle \text{, }\left( r>1\right) ,
\end{eqnarray}%
where $K=2n\pi /N,$ $n\in \left[ -N/2,N/2\right] $ is the momentum vector,
and $\pm $ denote two degenerate subspaces originating from the\textbf{\ }$%
\mathcal{P}$ symmetry. A two-particle eigenstate has the form\textbf{\ }of%
\begin{equation}
\left\vert \psi _{K,k}^{\pm }\right\rangle =\sum_{r}F_{K,k}^{\pm }\left(
r\right) \left\vert \varphi _{r}^{\pm }\left( K\right) \right\rangle \text{,}
\end{equation}%
with the condition $F_{K,k}^{\pm }\left( -1\right) =0$, where the two
degenerate wave functions $F_{K,k}^{\pm }\left( r\right) $\ satisfy the Schr%
\"{o}dinger equations%
\begin{eqnarray}
Q_{r}^{K}F_{K,k}^{\pm }\left( r+1\right) +Q_{r-1}^{K}F_{K,k}^{\pm }\left(
r-1\right) + &&  \notag \\
\lbrack \pm J\delta _{r,0}+\left( -1\right) ^{n}Q_{r}^{K}\delta
_{r,N_{0}}-\varepsilon _{K}]F_{K,k}^{\pm }\left( r\right) =0, &&  \label{SE}
\end{eqnarray}%
with the eigen energy $\varepsilon _{K}$\ in the invariant subspace indexed
by $K$. Here the factors are $Q_{r}^{K}=-2\sqrt{2}\kappa \cos \left(
K/2\right) $ for $r=0$ and $-2\kappa \cos \left( K/2\right) $ for $r\neq 0$,
respectively. It indicates that the eigen problem of two-particle matrix can
be reduced to\ a\ single-particle governed by the equivalent Hamiltonians%
\begin{equation}
H_{\text{eq}}^{K,\pm }=\pm J\left\vert 0\right\rangle \left\langle
0\right\vert +\sum_{i=1}^{\infty }\left( Q_{i}^{K}\left\vert i\right\rangle
\left\langle i+1\right\vert +\text{H.c.}\right) ,
\end{equation}%
which clearly represents a semi-infinite chain with the ending on-site
potential $J$. We are concerned with only the scattering solution by the $0$%
th end. The Bethe ansatz solutions have the form
\begin{equation}
F_{K,k}^{\pm }\left( r\right) =e^{-ikr}+R^{\pm }e^{ikr}.
\end{equation}%
Substituting $F_{K,k}^{\pm }\left( r\right) $\ into Eq. (\ref{SE}), we have
\begin{equation}
\varepsilon _{K}\left( k\right) =-4\kappa \cos \left( K/2\right) \cos k,k\in
\left[ 0,\pi \right] ,
\end{equation}%
and%
\begin{equation}
R_{K,k}^{\pm }=\frac{i\lambda _{K,k}\pm J}{i\lambda _{K,k}\mp J}=e^{\pm
i\Delta _{K,k}},  \label{fkj}
\end{equation}%
with%
\begin{eqnarray}
\lambda _{K,k} &=&4\kappa \cos \left( K/2\right) \sin k, \\
\Delta _{K,k} &=&2\tan ^{-1}\left( -\frac{J}{\lambda _{K,k}}\right) .
\end{eqnarray}%
For convenience in the application of wavepacket dynamics, we rewrite the
solutions in the form%
\begin{equation}
\left\vert \psi _{K,k}^{\pm }\right\rangle =\sum_{r,\rho }f_{K,k}^{\rho ,\pm
}\left( r\right) \left\vert \phi _{r}^{\rho }\left( K\right) \right\rangle ,
\end{equation}%
where $\rho =A,B$ and%
\begin{eqnarray}
&&\left\vert \phi _{0}^{\rho }\left( K\right) \right\rangle =\frac{1}{2\sqrt{%
N}}\sum_{j}e^{iKj}a_{\rho ,j}^{\dagger }a_{\rho ,j}^{\dagger }\left\vert
\text{vac}\right\rangle , \\
&&\left\vert \phi _{r}^{\rho }\left( K\right) \right\rangle =\frac{1}{\sqrt{N%
}}e^{iKr/2}\sum_{j}e^{iKj}a_{\rho ,j}^{\dagger }a_{\rho ,j+r}^{\dagger
}\left\vert \text{vac}\right\rangle \text{, }  \notag \\
&&\left( r>1\right) .
\end{eqnarray}%
The corresponding wavefunctions $f_{K,k}^{\rho ,\pm }\left( r\right) $ can
be expressed as
\begin{eqnarray}
f_{K,k}^{A,+}\left( r\right) &=&f_{K,k}^{B,-}\left( r\right)  \notag \\
&=&\left\{
\begin{array}{c}
e^{-ikr}+\frac{\lambda _{K,k}^{2}-J^{2}}{\lambda _{K,k}^{2}+J^{2}}e^{ikr},r>0
\\
\left( 1+\frac{\lambda _{K,k}^{2}-J^{2}}{\lambda _{K,k}^{2}+J^{2}}\right) /%
\sqrt{2},r=0%
\end{array}%
\right. ,
\end{eqnarray}%
and%
\begin{eqnarray}
f_{K,k}^{B,+}\left( r\right) &=&f_{K,k}^{A,-}\left( r\right)  \notag \\
&=&\left\{
\begin{array}{c}
-\frac{2i\lambda _{K,k}J}{\lambda _{K,k}^{2}+J^{2}}e^{ikr},r>0 \\
\frac{-\sqrt{2}i\lambda _{K,k}J}{\left( \lambda _{K,k}^{2}+J^{2}\right) },r=0%
\end{array}%
\right. .
\end{eqnarray}

\subsection{Solution of the non-Hermitian Hubbard model}

\label{sec_App_2}Similarly, considering the Hamiltonian $\mathcal{H}$, we
find that it admits all the symmetries we used for solving the eigen problem
of $H$. Then a two-particle state for $\mathcal{H}_{\rho }$\ is written as%
\begin{equation}
\left\vert \kappa _{K}^{\rho }\right\rangle =\sum_{r}G_{K,k}^{\rho }\left(
r\right) \left\vert \phi _{r}^{\rho }\left( K\right) \right\rangle \text{, }%
\left( G_{K,k}^{\rho }\left( -1\right) =0\right)  \label{eig}
\end{equation}%
where wave functions $G_{K,k}^{\rho }\left( r\right) $\ satisfy the Schr\"{o}%
dinger equations%
\begin{eqnarray}
Q_{r}^{K}G_{K,k}^{\rho }\left( r+1\right) +Q_{r-1}^{K}G_{K,k}^{\rho }\left(
r-1\right) + &&  \notag \\
\lbrack iU_{\rho }\delta _{r,0}+\left( -1\right) ^{n}Q_{r}^{K}\delta
_{r,N_{0}}-\epsilon _{K}]G_{K,k}^{\rho }\left( r\right) =0, &&
\end{eqnarray}%
with the eigen energy $\epsilon _{K}$\ in the invariant subspace indexed by $%
K$. We are concerned with only the scattering solution by the $0$th end. In
this sense, $G_{K,k}^{\rho }$ can be obtained from the two equivalent
Hamiltonians in two subspaces%
\begin{equation}
\mathcal{H}_{\text{eq}}^{K,\rho }=iU_{\rho }\left\vert 0\right\rangle
\left\langle 0\right\vert +\sum_{i=0}^{\infty }\left( Q_{i}^{K}\left\vert
i\right\rangle \left\langle i+1\right\vert +\text{H.c.}\right) .
\end{equation}%
By the same procedures, we have%
\begin{equation}
G_{K,k}^{\rho }\left( r\right) =\left\{
\begin{array}{c}
e^{-ikj}+\frac{\lambda _{K,k}+U_{\rho }}{\lambda _{K,k}-U_{\rho }}e^{ikj},r>0
\\
\left( 1+\frac{\lambda _{K,k}+U_{\rho }}{\lambda _{K,k}-U_{\rho }}\right) /%
\sqrt{2},r=0%
\end{array}%
\right. .
\end{equation}%
with eigen energy $\epsilon _{K}\left( k\right) =-4\kappa \cos \left(
K/2\right) \cos k$, $k\in \left[ 0,\pi \right] $. Furthemore, we can rewrite
the solution in the form%
\begin{equation}
g_{K,k}^{\pm }\left( r\right) =\left[ G_{K,k}^{A}\left( r\right) \pm
G_{K,k}^{B}\left( r\right) \right] /\sqrt{2}.
\end{equation}

\subsection{Spectral singularity of Hubbard chain}

\label{sec_App_3}We note that wave function $G_{K,k}^{\rho }\left( r\right) $%
\ only\ depends on $\rho $ via $U_{\rho }$. This is because the two chains $%
A $ and $B$ are independent. Then $G_{K,k}^{\rho }\left( r\right) $\
actually represents the two-particle solution of a non-Hermitian Hubbard
Hamiltonian on a single chain $\rho $\ with on-site imaginary\textbf{\ }%
interaction strength $iU_{\rho }$. We find that $G_{K,k}^{\rho }\left(
r\right) \rightarrow \infty $ as $U_{\rho }=\lambda _{K,k}$, which indicates
a spectral singularity at $\left\{ K,k\right\} $ \cite{ZXZ,AMSS}.

\acknowledgments X. Z. Zhang thanks S. J. Yuan for helpful discussions and
comments. This work is supported by the National Natural Science Foundation
of China (Grants No. 11505126 and No. 11374163). X. Z. Zhang is also
supported by the Postdoctoral Science Foundation of China (Grant No.
2016M591055) and PhD research start-up foundation of Tianjin Normal
University under Grant No. 52XB1415.

\end{document}